\documentstyle[aps]{revtex}
\begin{document}

\title{
New model for system of mesoscopic Josephson contacts.
}

\author{A.I. Belousov, Yu.E. Lozovik
\thanks{e-mail:lozovik@isan.troitsk.ru}}
\address{Институт спектроскопии РАН, 142092 Троицк, Московская обл., Россия.}
\maketitle

\begin{abstract}
Quantum fluctuations of the phases of the order parameter in two - dimensional
(2D) arrays of mesoscopic Josephson junctions and their effect on the
destruction of superconductivity in the system are investigated by means of
a quantum-cosine model that is free of the incorrect application of
the phase operator. The proposed model employs trigonometric
phase operators and makes it possible to study arrays of
small superconducting granules, pores filled with superfluid helium,
or Josephson junctions in which the average number of particles
$n_0$ (effective bosons, He atoms, and so on) is small, and the
standard approach employing the phase operator and the particle
number operator as conjugate ones is inapplicable. There is a large
difference in the phase diagrams between arrays of macroscopic
and mesoscopic objects for $n_0 < 5$ and $U<J$
($U$ is the characteristic interaction energy of the particle per
granule and $J$ is the Josephson coupling constant).
Reentrant superconductivity phenomena are discussed.
\end{abstract}

PACS: 73.23.Ps, 74.50.+r, 74.20.Mn

\vspace{0.5cm}

The development of microlithographic methods has now made it possible to
produce regular arrays of extremely small metallic granules, Josephson
junctions, and so on. The study of the properties of such objects is
of interest not only from the fundamental standpoint but also
in connection with their possible application in nanoelectronics.
for this reason, a great deal of attention is now being devoted to the
investigation of models reflecting the main properties of mesoscopic Josephson
arrays (see Refs \cite{Anderson} - \cite{Dobr} and the literature cited
therein).

In the present letter we call a granule (or a pore containing superfluid He)
{\it mesoscopic} if the rms fluctuation of the number of effective
bosons viz. "Cooper pairs" (or $He$ atoms in a pore),
are comparable to the average number of particles. We do not
address the question of the character of the transition of an individual
granule to the superconducting state (or helium in a pore to the
superfluid state), making the assumption that this transition
(or crossover) has already occurred at a higher temperature $T_{c_0}$
and that the concept, which we employ below, of an effective boson
(particle) is defined. In this connection we note only that,
just as for nucleon pairing in nuclei, strictly speaking,
there does not exist a critical size of granules in which
pairing is possible (but, for example, the parity of the number of electrons,
the character of the fillings of the levels,and so on are important).
The exposition below concerns an array of mesoscopic superconducting granules,
but the analysis can obliviously extended to the case of superfluid helium
in a porous media.

In the present letter we show that the application of the phase and
particle number operators as conjugate variables \cite{Anderson}
ordinarily employed for
describing such systems (in a quantum XY model) is limited to systems of
{\it macroscopic} granules, while in the case of a small
average number of particles per granule other models that do not
employ the incorrect "phase operator" are required (see below).
Recently, a boson Hubbard model that takes into account of not only quantum
fluctuations of the phase, but also the modulus of the superconducting
order parameters was used \cite{ShonPhysScri,Belous_JETP}
to investigate the superconducting properties
of an array of mesoscopic granules. However, it is of interest
to examine a system in which the fluctuations of the local
superfluid density on the granules are small even in the mesoscopic region,
and the quantum fluctuations of the phases of the order parameter
play the main role in the destruction of the global superconducting state
of the array.

At temperatures below the temperature $T_{c_0}$
at which superconductivity
is established in an individual granule the state of the system is determined
by two parameters \cite{Belous_JETP,Akopov}:
a) the ratio of the characteristic Coulomb interaction energy
$E_C \sim U/2 $ of the particles in a granule with self - capacitance
$C_0 = 4 e^2 / U$
and the Josephson energy for particles tunneling between granules
$E_J \sim J $,
the corresponding dimensionless quantum parameter being
$q \equiv \sqrt{U / J}$, and 2)
the dimensionless temperature of the system $T \equiv k_b T / J$.

\vspace{0.2cm}

{\bf 1.} \quad
The superconducting properties of an array of "macroscopic" superconducting
granules have traditionally been investigated in a quantum XY model with the
Hamiltonian
\begin{eqnarray}
\hat H_{XY} =
J \sum\limits_{<i,j>}
\left( 1 - \cos{(\hat \varphi_i - \hat \varphi_j)} \right) +
\frac{U}{2} \sum\limits_i
(\hat n_i - n_0)^2
\label{hamiltonian_xy}
\end{eqnarray}
where the sum $\sum_{<i,j>}$ in Eq.(\ref{hamiltonian_xy})
extends over all nonrepeating pairs $<i,j>$
of neighboring sites. The phases of the order parameter
$\varphi_i \in [0,2\pi)$
and the particle number operator $\hat n_i$ in the $i$th granule is assumed,
starting with Anderson's work\cite{Anderson} to be conjugate to the
"phase operator"
$\hat \varphi_i$: \quad
$\hat n_i - n_0 = \tilde \jmath \partial / \partial \varphi_i$.
We note that the direct application
of such a "phase operator" $\hat \varphi$
is, strictly speaking, incorrect, if
for no other reason than that its effect is to transfer
the state out of the domain of definition
(the set of periodic $\psi(\varphi + 2\pi) = \psi(\varphi)$).
Moreover, direct quantization of the order parameter as
$\psi = \Delta e^{\tilde \jmath \varphi} \to
\hat a = \sqrt{\hat n} \hat e^{\tilde \jmath \varphi}$
gives, in view of the boundness of the spectrum of the particle number
operator $\hat n = \hat a^{\dagger} \hat a$,
a nonunitary operator $\hat e^{\tilde \jmath \varphi}$.
This makes it impossible to introduce a hermitian phase operator and
leads to many paradoxes \cite{QPhase}.

Many such difficulties can be circumvented by considering the
"trigonometric" operators  $\widehat{cos} \; \varphi$
and $\widehat{sin} \; \varphi$ (which do not commute with each other),
while leaving the phase operator $\hat \varphi$ itself undefined
 \cite{Louisell}.
In this approach, all operators of physical quantities which are periodic
functions of the phase (in terms of the quantum XY model) must be
rewritten as sums of trigonometric functions, followed by the substitutions
$$
\cos{(\varphi)} \to \widehat{cos} \; \varphi =
\frac{1}{2}\left( \hat a^{\dagger} \frac{1}{\sqrt{\hat n + 1}}  +
\frac{1}{\sqrt{\hat n + 1}} \hat a \right)
$$
$$
\sin{(\varphi)} \to \widehat{sin} \; \varphi =
\frac{\tilde \jmath}{2}\left(
\hat a^{\dagger} \frac{1}{\sqrt{\hat n + 1}}  -
\frac{1}{\sqrt{\hat n +1}} \hat a \right)
$$
where the operators $\hat a^{\dagger}$ and $\hat a$ are Bose particle
creation and annihilation operators,
and $\hat n = \hat a^{\dagger} \hat a$.
As $n_0$  increases, when the rms fluctuations of
the particle number in a granule are much smaller than their average value,
the boundness of the spectrum of the operator $\hat n$  becomes unimportant
and the "quantum trigonometric operators"
$\widehat{cos} \; \varphi$ and $\widehat{sin} \; \varphi$
introduced above transform into their quasiclassical limit
$\cos{(\varphi)}$ and $\sin{(\varphi)}$.

Applying the procedure described above to the Hamiltonian
(\ref{hamiltonian_xy}) for our system,we have
\begin{eqnarray}
\hat H =
J \sum\limits_{<i,j>}
\left( 1 - \widehat{cos} \; \varphi_i  \widehat{cos} \; \varphi_j -
\widehat{sin} \; \varphi_i  \widehat{sin} \; \varphi_i \right) +
\frac{U}{2} \sum\limits_{i}
\left( \hat n_i - n_0 \right)^2
\label{hamiltonian}
\end{eqnarray}
In what follows we shall, for brevity, refer to the model (\ref{hamiltonian})
as the {\it quantum cosine model}.

In the present letter we are interested in the system (\ref{hamiltonian})
with {\it integral} occupation numbers, when the average number of bosons
in each granule $n_0 = \langle a^{\dagger}_i a_i \rangle$
is an {\it integer}. It can be shown
(see Ref.\cite{Cha} for an analysis of this case in the boson Hubbard model)
that under this condition and at $T=0$ the model (\ref{hamiltonian})
belongs to the same universality class as the quantum XY model
(\ref{hamiltonian_xy}).
However, at {\it finite} temperatures the critical behavior of the
quantum cosine model will be identical -- in contrast to
the system at zero temperature -- to the critical behavior of the quantum
XY model in some {\it finite} range of values of the average
occupation number $n_0$ of the granules in the system
(near integral values of $n_0$).

\vspace{0.2cm}

{\bf 2.} \quad
To calculate the properties of the system (\ref{hamiltonian})
in the plane of controlling
parameters $\{q,T\}$, we used the quantum Monte Carlo method of integration
along trajectories in a "checkerboard" modification\cite{Blaer},
where the degrees of
freedom of the discretized system are the occupation numbers $\{n_i^p\}$
of the sites of a three - dimensional lattice $N \times N \times 4P$
formed by $4P$-fold multiplication of the initial $N \times N$ lattice
along the imaginary - time axis.

In analyzing the superconducting properties of the array,
attention was focused mainly on the analysis of the superfluid
fraction (the analog of the "helicity modulus" in the quantum XY model;
see Ref. \cite{Minnhagen}), the expression for which in terms of the quantum
cosine model (\ref{hamiltonian}) has the form
\begin{eqnarray}
\label{helicity}
\nu_s = - \frac{1}{N^2}
\left< \hat T_x \right> - \frac{1}{N^2 T P}
\sum\limits_{\tau=0}^{P-1} \left< \hat J_x^{(p)}(\tau) \hat J_x^{(p)}(0) \right>
\\
\hat T_x = - \sum\limits_i \left(
\widehat{cos} \; \varphi_{i+x}  \widehat{cos} \; \varphi_i +
\widehat{sin} \; \varphi_{i+x}  \widehat{sin} \; \varphi_i
\right),
\nonumber
\\
\hat J_x^{(p)} = \sum\limits_i
\left(
\widehat{sin} \; \varphi_{i+x}  \widehat{cos} \; \varphi_i -
\widehat{cos} \; \varphi_{i+x}  \widehat{sin} \; \varphi_i
\right), \qquad
\hat J_x^{(p)}(\tau) = e^{\tau \beta \hat H / P} \hat J_x^{(p)}
e^{-\tau \beta \hat H / P}
\nonumber
\end{eqnarray}
The superfluid fraction was also found in terms of the fluctuation of the
"winding number" \cite{Trivedi,Blaer}.

An important quantity required for investigating the role of the mesoscopic
nature of the system are the fluctuations of the particle number over the
granule
\begin{eqnarray}
\delta n^2 = \frac{1}{4P N^2}
\left< \sum\limits_{p=0}^{4P-1}\sum\limits_{i} (n_{i}^p - n_0)^2 \right>
\label{n_fluct}
\end{eqnarray}

\vspace{0.2cm}

{\bf 3.} \quad
Let us examine the Monte Carlo results for the quantum cosine model
(\ref{hamiltonian}).
The main quantity describing the state of the system at a given point
$\{q,T\}$ of the phase diagram (Fig.~1) is the superfluid fraction $\nu_s$.
Vanishing of $\nu_s$ indicates disordering of the system. The measured curves
of the superfluid fraction versus the temperature $T$ for fixed values of
quantum parameter $q$ are presented in Fig.~2. This figure shows the
computational results for a $N \times N \ 6 \times 6$ system with
$n_0 = 1,3,5,7$ and, for comparison, for the quantum XY model
(\ref{hamiltonian_xy}).
For a weak interparticle interaction in the granules $q<1$
(which in terms of the XY model corresponds to a small role of quantum
fluctuations of the phase of the order parameter) the superconductor --
metal transition temperature $T_c$ depends strongly on the average
number of particles $n_0$ in a granule (see Fig.~2a).
The temperature $T_c(q)$ can be estimated from the universal
relation $\nu_s(q;T_c) = 2T_c / \pi$ \cite{Minnhagen}. It is
evident from the figure that to suppress fluctuation phenomena in a system
of mesoscopic granules or pores requires {\it lower} temperatures
than in the case of systems of macroscopic granules.
As the particle density increases (transition to the case of an
array of macroscopic granules), for $n_0>5$ the plots
of $\left.\nu_s(T;n_0)\right|_{q=const}$ merge, to within
the limits of the measurement error, and the model (\ref{hamiltonian})
goes over to its limit -- the quantum XY model.
This observation is also confirmed in Fig.~3 which displays the dependence
of the superfluid fraction in models (\ref{hamiltonian}) and
(\ref{hamiltonian_xy}) on  the quantum parameter
$q$ at $T=0.5$. We note that for $q>1$ the properties of the models
under study differ very little, and therefore the effect
of the mesoscopicity of the granules in the array are very small.
This is confirmed in Figure~2b.

An interesting effect which we found in our calculation is
{\it re - entrant superconductivity} of an array of mesoscopic
granules with respect to the parameter $q$
(determining the characteristic particle interaction energy).
The {\it increase} in the density of the superfluid component
with increasing quantum parameter $q$ at a fixed temperature $T$,
clearly seen in Fig.~3, is confirmed by computational results for the model
(\ref{hamiltonian}) in mean field theory \cite{Verzakov} (see Fig.~1).

The phase diagram of an ordered two-dimensional Josephson array of
mesoscopic granules was constructed using the results presented
above (see Fig.~1). For comparison, the computational results obtained
in mean field theory, which is in qualitative agreement
with the Monte Carlo data, is also shown. We note, for comparison, that
taking the fluctuations of the modulus of the order parameter
(local superfluid density) into account in the boson Hubbard model
{\it increases} the superconducting transition temperature of a
mesoscopic system \cite{Belous_JETP}.

The character of the phase transition occurring along the line $T_c(q)$
can be analyzed in greater detail by studying the fluctuations
of the particle number over the granules (\ref{n_fluct})
as a function of the temperature and quantum parameter.
The calculations show that as the temperature increases,
the fluctuations of the particle number over the granules increases,
as is characteristic for a transition to a state with a higher
conductivity. Therefore, at finite temperatures the line of phase transitions
$T_c(q)$ (see Fig.~1) is a line of superconductor - to - metal
transitions. Conversely, at a fixed temperature the particle- number
fluctuations $\delta n^2(q)$ as a function of the quantum parameter $q$
decrease, as is characteristic for a (finite - temperature crossover)
transition to a Mott insulator state \cite{Girvin}.

From the relative rms particle - number fluctuations
$\epsilon_n \equiv \sqrt{\delta n^2 / n_0^2}$ it can be concluded
that the value $\epsilon = \epsilon_n^{mes} \sim 0.5$
can be regarded as a criterion for determining the mesoscopicity
of granules in the array, namely, for smaller relative fluctuations
the system can be viewed as consisting of macroscopic granules
and can be described by the quantum XY model (\ref{hamiltonian_xy}).

\vspace{0.2cm}

{\bf 4.} \quad
Thus, a new model has been proposed for systems of mesoscopic Josephson
junctions - the quantum cosine model (\ref{hamiltonian}),
which takes into account the quantum
fluctuations of the phases of the order parameter and does {\it not}
employ an incorrect definition of the "phase operator". This model
can be used to investigate the properties of systems of mesoscopic
granules or pores containing a superfluid $He$, where the relative
fluctuation of the "effective bosons" over the granules or
atoms of the liquid in the pores are large and the quantum XY model
(\ref{hamiltonian_xy}) is inapplicable.

The computational results show that in the case of a system of weakly
interacting particles (for $q<1$) the temperature at which a global
superconducting state appears depends strongly on the particle density,
approaching {\it from below} the metal - superconductor transition
temperature in a system of macroscopic granules or pores. It was found
that the corresponding "macroscopic" limit, where the system
is adequately described in a quantum XY model
(\ref{hamiltonian_xy}), is reached at
comparatively low densities, $n_0 \sim 5$.

In the region of large quantum fluctuations of the phases ($q > 1$),
the relative fluctuations of the particle number over the granules
are strongly suppressed by the interaction, and mesoscopic effects are
important only at low temperatures ($T < 0.5$) and low
densities ($n_0 \sim 1$).

As one can see from the results of calculations, the proposed quantum cosine
model (\ref{hamiltonian}) does not (at least in the investigated range of the controlling
parameters) exhibit re-entrant superconductivity with respect to
{\it temperature}, where for some values of the quantum parameter $q$
a global superconducting state is absent at both high and low temperatures.
However, for a weak interparticle interaction , when $q<1$, there is
re-entrant superconductivity with respect to the {\it quantum parameter}
$q$. We found that for the model (\ref{hamiltonian_xy}),
in contrast to the behavior of the
Hubbard model and the quantum XY model, the degree of disorder
in the system increases with increasing interaction of the bosons
(with decreasing strength of quantum fluctuations of the phases
in terms of the quantum XY model).

\vspace{0.3cm}

We thank V.F.~Gantmakher for a helpful discussion of the results of this
work. This work was supported by grants from the Russian Fund for
Fundamental Research and the programs "Physics of Solid-State Nanostructures"
and "Surface Atomic Structures".

\vspace{0.2cm}

Fig.~1   \\
Phase diagrams of the quantum, cosine model (QC),
the boson Hubbard model (H) \cite{Belous_JETP}, and
the quantum XY model (2+1 XY) \cite{Belous_FTT}.
S - superconducting state, N - normal state.
\\
Insert: Mean field theory results \cite{Verzakov}:
{\bf 1} - $n_0 = 1$;  {\bf 2} - $n_0 = 2$; {\bf 3} - $n_0 = 6$;
{\bf 4} - quantum XY model ($n_0 = \infty$); \\
Here and below the symbols represent the quantum Monte Carlo results:
open symbols - $N=10$, filled symbols - $N=6$,
squares - $n_0=1$, circles - $n_0=3$, triangles - $n_0=5$,
rhombi - $n_0=7$, asterisks - quantum XY model.

\vspace{0.2cm}

Fig.~2  \\
Superfluid fraction $\nu_s$ (helicity modulus $\gamma$ in the case of
quantum XY model) versus temperature $T$:
{\bf a)} $q=0.2$; {\bf б)} $q=2.0$;
The dotted line shows the curve $2T/\pi$
A spline interpolation is drawn in as an aid to the eye. The statistical
errors, which are not shown, are less than the size of the corresponding
symbols.

\vspace{0.2cm}

Fig.~3  \\
Superfluid fraction $\nu_s$ (helicity modulus $\gamma$ in the case of
quantum XY model) versus the quantum parameter $q$ at $T=0.5$;
the dotted line shows the line $1/\pi$. Insert: $\nu_s(q)$ at $T=1$.

\vspace{0.2cm}

Translated by M.E.~Alferieff.
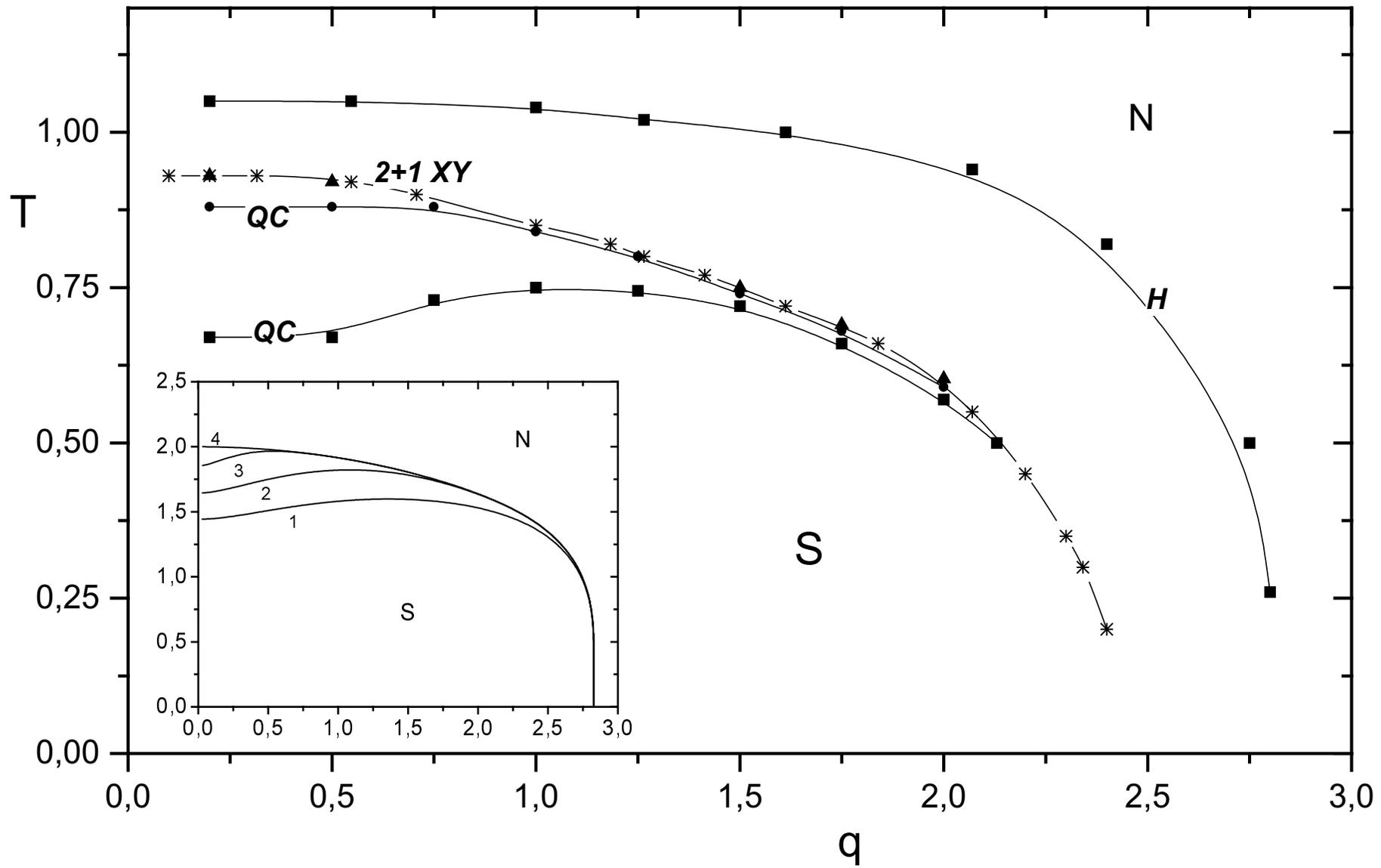
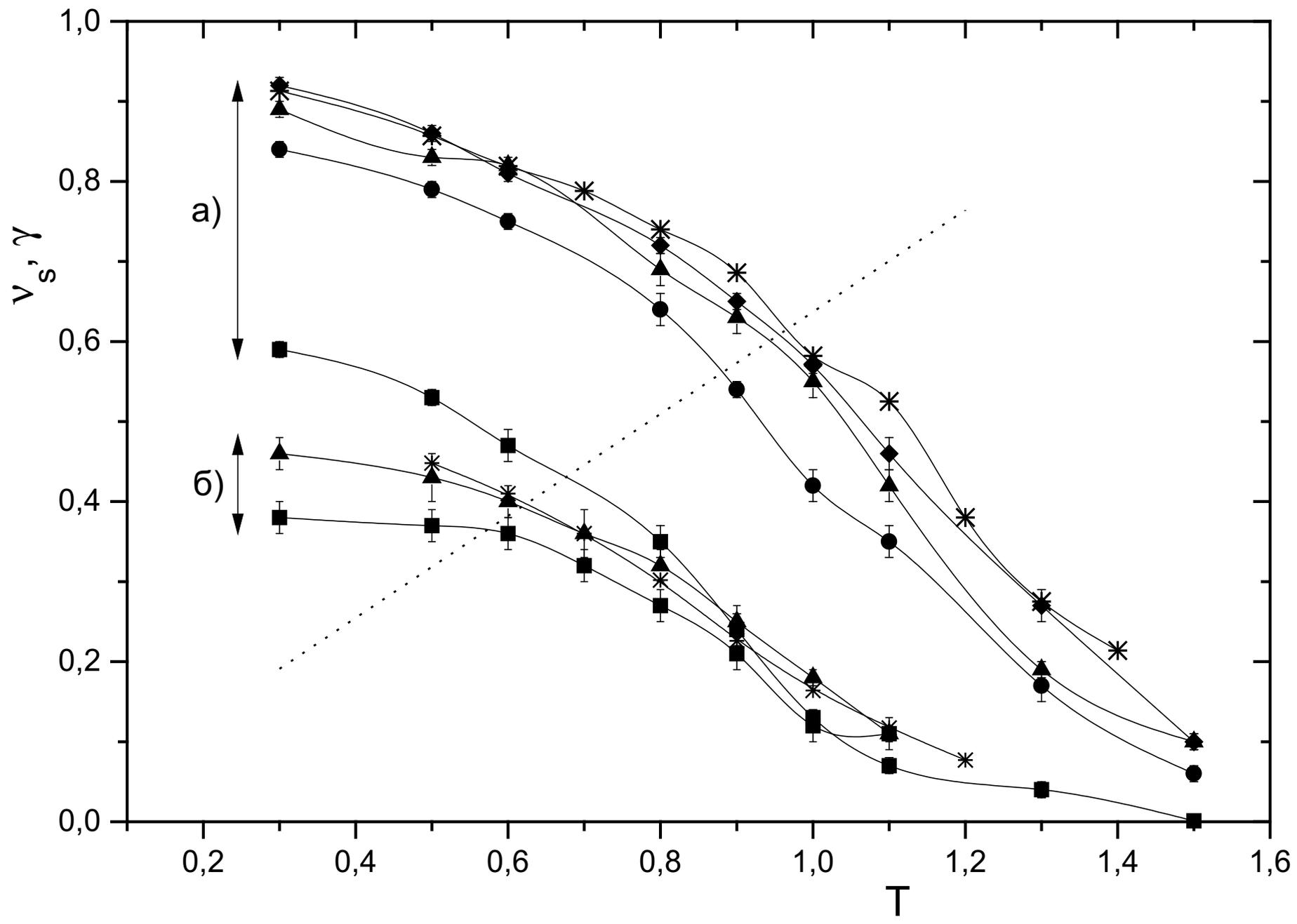
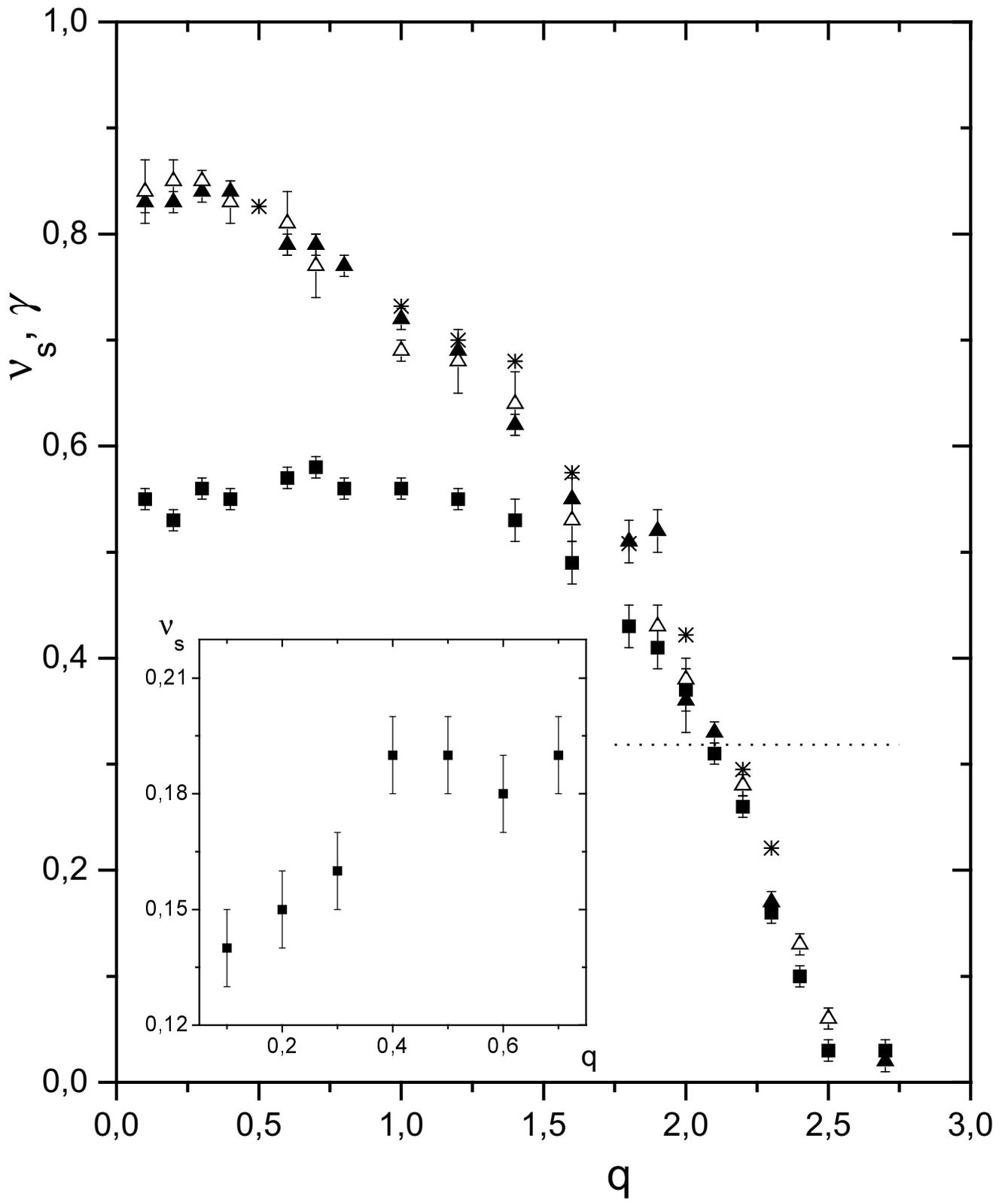
\end{document}